# Automation and miniaturization of Golden Gate DNA assembly reactions using acoustic dispensers

Tania S. Köbel[1], Daniel Schindler[1,2,*]

[1] Max Planck Institute for Terrestrial Microbiology, Karl-von-Frisch-Str. 10, D-35043 Marburg, Germany

[2] Center for Synthetic Microbiology (SYNMIKRO), Philipps-University Marburg, Karl-von-Frisch-Str. 14, D-35032 Marburg, Germany

* Correspondence to: daniel.schindler@mpi-marburg.mpg.de

**Abstract**
Golden Gate cloning has become one of the most popular DNA assembly techniques. Its modular and hierarchical structure allows the construction of complex DNA fragments. Over time, Golden Gate cloning allows for the creation of a repository of reusable parts, reducing the cost of frequent sequence validation. However, as the number of reactions and fragments increases, so does the cost of consumables and the potential for human error. Typically, Golden Gate reactions are performed in volumes of 10 to 25 µL. Recent technological advances have led to the development of liquid handling robots that use sound to transfer liquids in the nL range from a source plate to a target plate. These acoustic dispensers have become particularly popular in the field of synthetic biology. The use of this technology allows miniaturization and parallelization of molecular reactions in a tip-free manner, making it sustainable by reducing plastic waste and reagent usage. Here, we provide a step-by-step protocol for performing and parallelizing Golden Gate cloning reactions in 1 µL total volume.

## 1. Introduction

Golden Gate cloning has become one of the most popular DNA assembly strategies. Initially developed as a one-step, one-pot reaction, it quickly demonstrated its modular and high-throughput capabilities [1,2]. All Golden Gate technologies are based on the use of type IIS enzymes, which recognize a specific, directed motif and cleave any sequence at a defined distance. This principle allowed the development of Golden Gate cloning and its subsequent hierarchical modularization in the form of the Modular Cloning System (MoClo). Since its initial development, several organism-specific Golden Gate toolboxes have been developed for rapid and reliable engineering of the chosen host [3-6]. In addition, Golden Gate cloning has been used to standardize and/or modularize cloning steps for technologies, such as the rapid and modular construction of CRISPR/Cas gRNA arrays [7], genetic engineering toolboxes [8-10], post-transcriptional control toolkits [11,12], and synthetic chromosome construction [13,14].

A common feature of standard Golden Gate protocols is that the reactions are typically performed in 10 to 25 µl volumes. Modern scientific approaches often require an increased amount of DNA synthesis and assembly for the construction of large metabolic pathways or whole synthetic chromosomes [15,16]. In addition to the increasing number of DNA assemblies, the complexity of Golden Gate reactions is also increasing, with > 50 DNA parts being successfully assembled in a single step [17]. The standardization achieved by MoClo is further enabling the increased implementation of laboratory automation and is driving the establishment of biofoundries, dedicated sites to bundle technology for the automation of biological workflows. Acoustic dispensers have become very valuable in this context. Acoustic dispensers use sound to transfer liquid from a source plate to a target plate in fixed nanoliter-sized droplets (Fig. 1) [18]. This technology has recently been established for DNA assembly at the nL scale [19]. We use this technology on a daily basis for various applications, such as the Golden Gate based construction of synthetic small regulatory RNAs (sRNAs) for post-transcriptional control in bacteria [11].

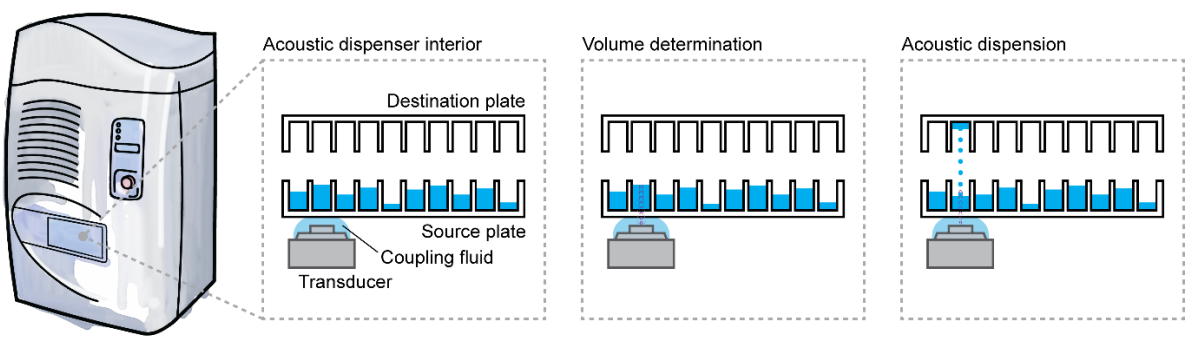

**Figure 1 | Principle and application of acoustic dispensers for molecular reactions.** Acoustic dispensers are systems controlled by an external computer. Within the acoustic dispenser there are three core elements: the transducer, the source plate, and the destination plate. The transducer is the sound source coupled to the source plate by the coupling fluid. The source plate contains the liquids to be dispensed into the target plate. The emitted sound waves are used to determine the volume of each individual well to confirm that an adequate volume is available at the

start of acoustic dispensing and to confirm volume reduction after dispensing. For acoustic dispensing, the sound energy is adjusted to eject machine-defined nL-sized droplets (e.g., Echo650T = 2.5 nL and Echo525 = 25 nL) from the source plate to the target plate.

Here, we provide a detailed step-by-step protocol for the use of acoustic dispensers for Golden Gate assemblies, their subsequent high-throughput transformation, and hit identification for subsequent validation and application. As a proof of concept, we have assembled transcription units for the expression of fluorescent proteins with unknown promoter sequences to identify promoters of varying strength for future applications. The process can be completed within 3-4 days, allowing for rapid and sustainable DNA assembly.

## 2. Materials

The following non-standard and standard laboratory equipment and supplies are required to perform this protocol.
1. Acoustic dispenser with cherry pick software and appropriate consumables
2. Standard PCR reagents, consumables and instrumentation
3. Standard reagents and equipment for gel electrophoresis
4. Standard reagents, consumables and instrumentation for microbial cultivation
5. Standard reagents, consumables and instrumentation for *E. coli* transformation
6. Standard micropipettes and consumables, 12-channel pipettes are recommended
7. A colony picking robot and a 96/384-pin based screening robot are recommended but not required

2.1 Plasmids
All plasmids relevant to this protocol are listed in Table 1. Plasmids can be obtained from the corresponding author. Any other Golden Gate cloning acceptor plasmid can be used. However, it is imperative to adapt the outlined protocol to the cloning standard with respect to the desired type IIS enzymes, corresponding overhangs and plasmid characteristics (e.g. antibiotic resistance) (*see* **Note 1, 2**).

**Tab. 1 | Plasmids used in this study.**

| Name | Relevant features | Parental plasmid | Reference |
|---|---|---|---|
| pSL009 | pBAD-TOPO derivative serving as empty plasmid control; Kan$^R$ | | [11] |
| pSL099 | Level 0 plasmid for subcloning of fragments to be released with SapI; Spec$^R$ | pMA60 [9] | [20] |
| pSL135 | Level 0 plasmid containing P$_L$lacO-1 promoter to be released with SapI; Spec$^R$ | pSL099 | [20] |
| pSL137 | pBAD-TOPO derivative allowing Golden Gate cloning of TUs with SapI; Kan$^R$ | pSL009 | [11] |
| pSL206 | Level 0 plasmid containing *mCherry* to be released with SapI; Spec$^R$ | pSL099 | this study |
| pSL269 | Level 0 plasmid containing a transcription terminator to be released with SapI; Spec$^R$ | pSL099 | this study |
| pSL541 | Level 0 plasmid containing *gfp* to be released with SapI; Spec$^R$ | pSL099 | this study |
| pSLcol_16.01 to pSLcol_16.48 | Level 0 plasmid collection containing distinct promoter sequences of a sequence library based on SLo5556 (*cf.* Fig. 2). Parts are released with SapI; Spec$^R$ | pSL099 | this study |

2.2 DNA oligonucleotides

The oligonucleotides used in the protocol are listed in Table 2. Oligonucleotides are dissolved in ddH$_2$O to make 100 µM stocks and stored at -20 °C. For PCR and sequencing reactions, 10 µM working stocks are generated with ddH$_2$O and stored at -20 °C.

**Tab. 2 | Oligonucleotides for the conduction of the described protocol.**

| ID | Sequence (5´-3´) | Information |
| --- | --- | --- |
| SLo1475 | `GCCTGACGAAGACCATTTC` | Forward primer to amplify sequence library |
| SLo1476 | `ACCTAGCCGAAGACCAACAT` | Reverse primer to amplify sequence library |
| SLo3961 | `CTGTCAAATGGACGAAGCAG` | Colony PCR primer to validate TU assembly |
| SLo3962 | `CAGGCAAATTCTGTTTTATCAGACC` | Colony PCR primer to validate TU assembly |
| SLo5556 | `CCTGACGAAGACCATTTCNNNNNNNNNNNNNNNNNT TGACANNNNNNNNNNNNNNNNGATACTNNNNNNNN AAAGAGGAGAAANNNNNATGTTGGTCTTCGGCTAG` | Degenerated sequence of a promoter serving as PCR template with ribosome binding site |

2.3 Enzymes

For this protocol, all enzymes used were provided by New England Biolabs (NEB) (*see* **Note 3**). However, enzymes from any supplier should be suitable for the protocol outlined.

1. Necessary type IIS enzyme(s), in this protocol SapI
2. T4 DNA Ligase (400,000 units/mL)
3. Taq DNA Polymerase (here, 2x ready-to-load Master Mix)

2.4 Antibiotics

All antibiotics used in this study are dissolved in ddH$_2$O and are stored in 1 mL aliquots at -20 °C. The stock concentration is 1,000 x.

1. Kanamycin (50 mg/mL stock)
2. Spectinomycin (120 mg/mL stock)

2.5 Chemicals, buffers, and media components

1. LB medium: 1% (w/v) tryptone, 0.5% (w/v) yeast extract, 1% (w/v) sodium chloride, pH 7.0 ± 0.2
2. 1 x TAE buffer: 1 mM EDTA · Na$_2$ · 2 H$_2$O, 20 mM acetate, 40 mM Tris
3. DNA dye: Thiazole orange dissolved in dimethyl sulfoxide (DMSO) (10,000 x stock concentration: 13 mg/mL)
4. Agarose (standard)
5. DNA ladder
6. Solid media is prepared with 2% Agar

2.6 Consumables
1. Single-well microtiter plates (*see* **Note 4**)
2. 120 x 120 mm square petri dishes (*see* **Note 5**)
3. 384-well PCR plates (*see* **Note 6**)
4. Adhesive PCR plate seal

2.7 Strains

All relevant laboratory *E. coli* strains for the outlined protocol are listed in Table 3. Cells were made competent using an in-house RbCl procedure [21] (*see* **Note 7**).

**Tab. 3 | Strains used in this study.**

| Name | Relevant features | Reference |
| --- | --- | --- |
| *E. coli* MG1655 | K-12 F$^-$ λ$^-$ (*see* **Note 8**) | [22] |
| *E. coli* DB3.1 | F$^-$ *gyrA462 endA1 glnV44* Δ(*sr1-recA*) *mcrB mrr hsdS20*(r$_B^-$, m$_B^-$) *ara14 galK2 lacY1 proA2 rpsL20*(Str$^R$) *xyl5* Δ*leu mtl1* (*see* **Note 9**) | Invitrogen |
| *E. coli* Top10 | F$^−$*mcrA* Δ(*mrr-hsdRMS-mcrBC*) φ80*lacZ*ΔM15 Δ*lacX74 nupG recA1 araD139* Δ(*ara-leu*)7697 *galE15 galK16 rpsL*(Str$^R$) *endA1* λ$^-$ (*see* **Note 10**) | Invitrogen |

## 3. Methods

### *3.1 Preparation of DNA for Golden Gate reaction*

The first step in performing Golden Gate experiments is to generate and purify DNA fragments to be assembled into the selected acceptor vector (Fig. 2A). The number of parts and cloning steps depends on the user's experimental design and the selected Golden Gate assembly standard. In this example, three level 0 parts are assembled into one transcription unit (TU) in a pBAD-derived acceptor plasmid. The focus of this chapter is on experimental design and the necessary steps to consider for highly parallelized, automated protocol execution. For this reason, we use a collection of promoter elements derived from a randomized sequence and fluorescent proteins as a direct readout of cloning success (Fig. 2B-C). The steps in Section 3.1 can be performed in about half a working day (excluding *E. coli* cultivation).

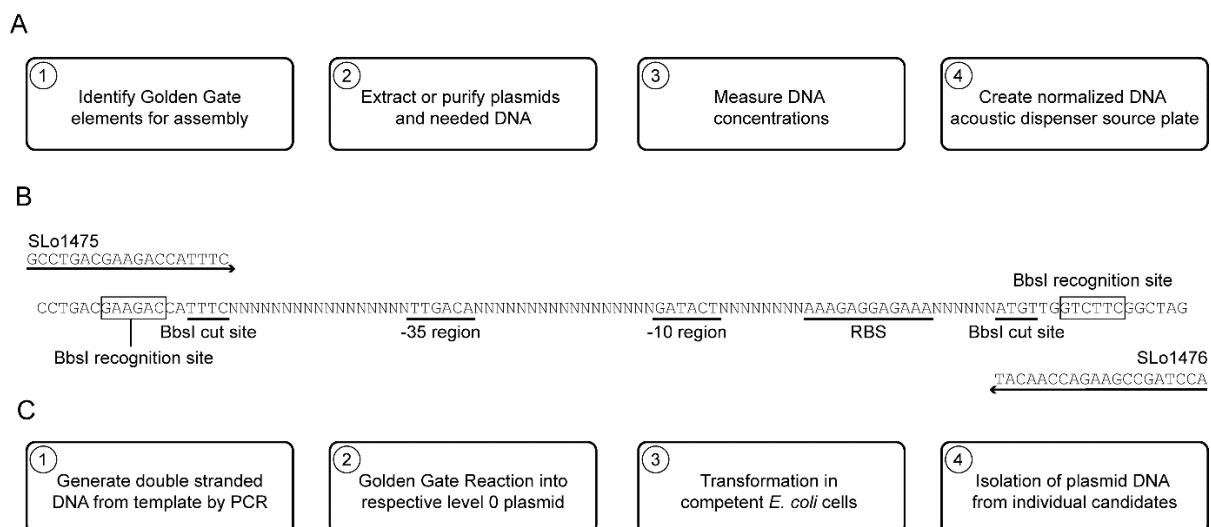

**Figure 2 | Workflow for DNA preparation for Golden Gate reactions. (A)** Outline of the steps required to perform acoustic dispenser-based Golden Gate assembly. The respective DNA parts to be assembled must be identified from the collection (or cloned and validated), purified, and the concentration must be measured before preparing the acoustic dispenser compatible source plate with normalized DNA concentrations. **(B)** Visualization of the promoter library concept. A degenerate oligonucleotide with defined -35, -10 region and ribosomal binding site (RBS) containing library amplification regions with BbsI restriction sites for Golden Gate cloning was designed and ordered. The indicated oligonucleotides (SLo1475/1476) were used to generate double-stranded DNA for Golden Gate cloning. The cloning of this element is not part of the protocol, but serves to clarify the promoter library used in this chapter. **(C)** Workflow for obtaining individual promoters from a random promoter sequence library.

1. Design the appropriate Golden Gate assembly and ensure that all parts and the appropriate acceptor vector are available. In this example, different promoters of unknown sequence from isolates of a complex library (pSLcol_16.01 to pSLcol_16.48) are individually combined with one of the two fluorophores (pSL206 or pSL541) and a terminator (pSL269) into the expression plasmid pSL137 (Fig. 2B-C). The expression plasmid is a pBAD derivative for direct use in *E. coli* cells. The experimental setup results in 96 Golden Gate DNA assemblies by generating all combinations of the 48 promoter sequences with each of the two fluorophores.
2. Perform plasmid extraction of the selected level 0 parts and the acceptor plasmid(s) (*see* **Note 11**). If PCR fragments with appropriate Golden Gate compatible overhangs are used, perform the PCR purification of choice. Alternatively, oligonucleotides can be annealed and used directly at the appropriate dilution (*see* **Note 12**).
3. Quantify the DNA by the method of choice (*see* **Note 13**).
4. To simplify subsequent DNA assembly steps, set the concentration of the DNA parts to 20 fmol/µL, DNA parts are used at equimolar concentrations (2 fmol each) in Golden Gate DNA assemblies (*see* **Note 14**). Normalization is ideally performed in acoustic dispenser compatible source plate(s) in combination with a bulk dispenser to quickly adjust the DNA concentration for each well. The plates are carefully sealed with an appropriate seal for storage at -20 °C (*see* **Note 15**).

5. Carefully document the location and information of each DNA for each well for future use and possible troubleshooting.
6. Store the normalized DNA at -20 °C until use (*see* **Note 16**).

### *3.2 Golden Gate reaction using an acoustic dispenser*

Acoustic dispensers use sound to transfer nL droplets from a source plate to a destination plate. This technology allows reaction volumes to be drastically reduced and reactions to be parallelized. For molecular DNA assembly methods, a final volume of 1 µL in 384-well destination plates has been proven successful [19,11]. Section 3.2 requires little hands-on time, but the Golden Gate reaction can be the rate-limiting step and is preferably performed overnight (Fig. 3). Steps 3.1 and 3.2 together can be performed during a standard working day.

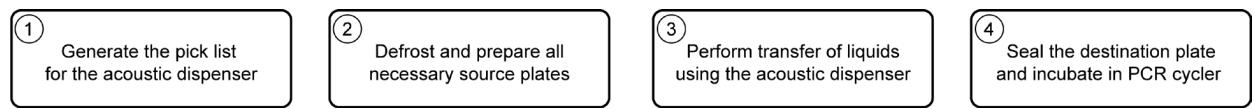

**Figure 3 | Workflow for Golden Gate reactions using an acoustic dispenser.** The first step is to create the pick list to define the appropriate acoustic dispenser liquid transfers. Ensure that all source plates are thawed and prepared to be used for the acoustic fluid transfer. After the transfer, seal the destination plate and perform Golden Gate DNA assembly in a PCR cycler.

1. Familiarize yourself with the acoustic dispenser available. Different dispensers have different specifications in terms of volume dispensed and liquid types. Consumables may vary from machine to machine, so it is imperative to be aware of all details before conducting experiments. Acoustic dispensers require regular maintenance. Important: Never use a machine without instruction from the local responsible operator.
2. Identify the appropriate source plate(s) for the Golden Gate assembly. For simplicity, this protocol requires one source plate containing all DNA parts as well as the reaction buffer, ddH$_2$O, and the enzymes (*see* **Note 17**). Prepare the pick list for the cherry pick software package (*see* **Note 18**). Below is an example pick list for assembling the first 12 TUs with *mCherry* and *gfp* for the outline example experiment. All DNA parts are used in equimolar concentrations:

Pick list (*see* **Note 19**):

| Source plate name | Source plate type | Source well | Destination plate name | Destination well * | Transfer volume | Part name |
|---|---|---|---|---|---|---|
| 1 | 384PP_Plus_AQ_SP | E1 | 1 | A1;A3;A5; […] E19;E21;E23 | 100 | T4 ligase buffer |
| 1 | 384PP_Plus_AQ_GP | E2 | 1 | A1;A3;A5; […] E19;E21;E23 | 100 | T4 ligase |
| 1 | 384PP_Plus_AQ_GP | E3 | 1 | A1;A3;A5; […] E19;E21;E23 | 100 | SapI |
| 1 | 384PP_Plus_AQ_SP | E4 | 1 | A1;A3;A5; […] E19;E21;E23 | 100 | pSL137 |
| 1 | 384PP_Plus_AQ_SP | E5 | 1 | A1;A3;A5; […] E19;E21;E23 | 100 | pSL269 |
| 1 | 384PP_Plus_AQ_SP | E6 | 1 | A1;A3;A5; […] A19;A21;A23 | 100 | pSL206 |
| 1 | 384PP_Plus_AQ_SP | E7 | 1 | E1;E3;E5; […] E19;E21;E23 | 100 | pSL541 |
| 1 | 384PP_Plus_AQ_SP | E8 | 1 | A1;A3;A5; […] E19;E21;E23 | 300 | ddH$_2$O |
| 1 | 384PP_Plus_AQ_SP | A1 | 1 | A1;E1 | 100 | pSLcol_16.01 |
| 1 | 384PP_Plus_AQ_SP | A3 | 1 | A3;E3 | 100 | pSLcol_16.02 |
| 1 | 384PP_Plus_AQ_SP | A5 | 1 | A5;E5 | 100 | pSLcol_16.03 |
| 1 | 384PP_Plus_AQ_SP | A7 | 1 | A7;E7 | 100 | pSLcol_16.04 |
| 1 | 384PP_Plus_AQ_SP | A9 | 1 | A9;E9 | 100 | pSLcol_16.05 |
| 1 | 384PP_Plus_AQ_SP | A11 | 1 | A11;E11 | 100 | pSLcol_16.06 |
| 1 | 384PP_Plus_AQ_SP | A13 | 1 | A13;E13 | 100 | pSLcol_16.07 |
| 1 | 384PP_Plus_AQ_SP | A15 | 1 | A15;E15 | 100 | pSLcol_16.08 |
| 1 | 384PP_Plus_AQ_SP | A17 | 1 | A17;E17 | 100 | pSLcol_16.09 |
| 1 | 384PP_Plus_AQ_SP | A19 | 1 | A19;E19 | 100 | pSLcol_16.10 |
| 1 | 384PP_Plus_AQ_SP | A21 | 1 | A21;E21 | 100 | pSLcol_16.11 |
| 1 | 384PP_Plus_AQ_SP | A23 | 1 | A23;E23 | 100 | pSLcol_16.12 |

\* […] indicates that A7 through A23 and E1 through E17 are omitted for visualization purposes.

3. Transfer the source plate from its storage to ice and ensure that all reagents are completely thawn (*see* **Note 20**). Allow the source plate equilibrate to room temperature prior to use to ensure proper transfer.
4. Centrifuge the source plate at max. 2000 x *g* for ~10 s to remove any air bubbles and to prevent a meniscus at the air/liquid interface (*see* **Note 21**).
5. Transfer the .csv file to the computer running the acoustic dispenser. Important: Follow your local IT best practice guidelines for data transfer.

6. Launch the cherry pick software to operate the acoustic dispenser. The instrument should automatically connect at this stage (*see* **Note 22**).
7. Open a new file in the cherry pick software to create a new protocol.
8. Adjust the settings in the category "protocol" according to the used equipment and consumables. In this chapter, the settings were as follows:
   Sample Plate Format:     384 PP
   Sample Plate Type:       384 PP_AQ_SP (for Echo525)
                            384 PP_AQ_SP2 (for Echo650T)
   Destination Plate Type:  384 PCR plate (*see* **Note 23**)
9. Navigate to the option "pick list" option and import the pick list created (*cf.* 3.2.2).
10. Double-check the selections, then click on the blue play button to save the experimental settings and to run the protocol.
11. Click on the start button and follow the instructions given by the operating system. Each step must to be confirmed individually by clicking the "ok" button. The steps are described below:
    a. Insert source plate
    b. Insert destination plate

12. The transfer is performed automatically based on the pick list provided. The system measures the volume of the source plate at the beginning and end of the experiment. The instrument provides a corresponding report file. It is recommended to always check the report file for "exception errors" (*see* **Note 24**).
13. After transfer, the source and destination plate are ejected and need to be handled accordingly.
14. After transfer, the source and destination plate are ejected and need to be handled accordingly.
15. Seal the source plate and return it to the appropriate storage conditions if not needed.
16. Seal the destination plate and transfer the reactions to a 384-well PCR cycler. Start the preferred Golden Gate assembly reaction protocol. This chapter uses a cycling program with the following settings for the assembly (*see* **Note 25**):

    [4 min 37 °C - 4 min 16 °C]$_{25\ cycles}$ - 10 min 50 °C - 20 min 65 °C - hold at 8 °C

17. Ideally, the Golden Gate reaction should be transformed to the preferred cells immediately after the reaction (see procedure in 3.3). However, Golden Gate reactions can be stored for a short time (< 48 hrs) in the refrigerator or for a longer time (> 48 hrs) at -20 °C.

### 3.3    *High-throughput transformation*
Golden Gate assemblies have proven to be very efficient. An advantage of using sequence verified level 0 parts is that there is usually no need to verify sequence integrity by sequencing in higher order assemblies. Depending on the complexity of the Golden Gate assembly, the user can usually estimate the percentage of successful assemblies. In most cases, picking two to four colonies is sufficient to obtain at least one successfully assembled construct (*see* **Note 26**). A limiting step is the high-throughput transformation of Golden Gate reactions into recipient cells. We use a workflow that allows high-throughput transformation into *E. coli* cells and subsequent selection of 12 or up to 96 transformations on a single agar plate. Step 3.3 requires little hands-on time and can be performed flexibly during a normal working day (Fig. 4). Selection plates are typically incubated overnight. The protocol uses chemically competent cells prepared in-house using the RbCl method [21]. However, any other method that produces sufficiently competent cells is applicable and is therefore not described in this protocol.

A
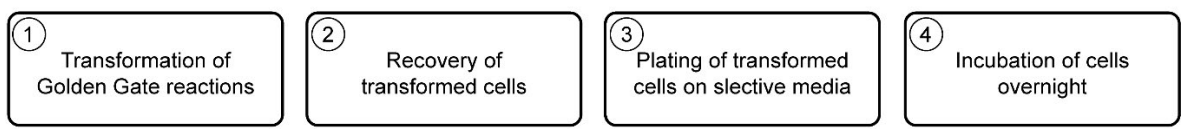

B  12-well drop gravity flow plating
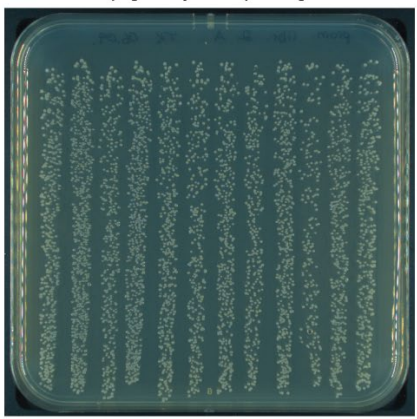
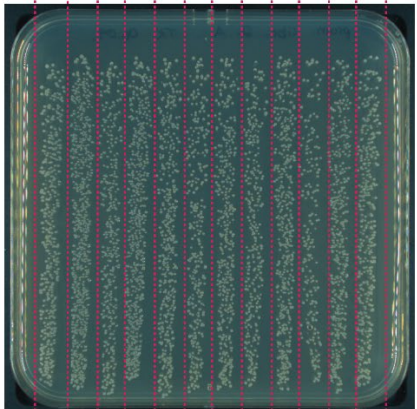

C  96-well transformation pinning
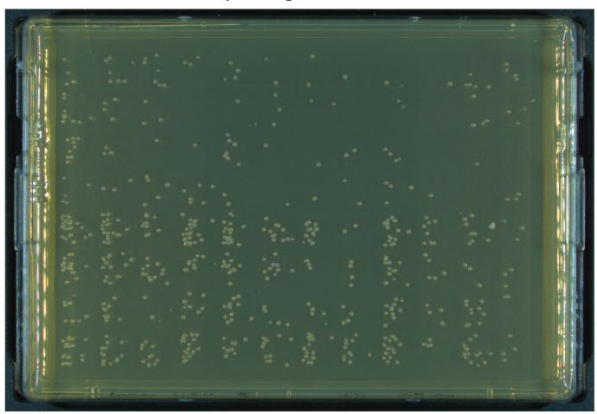
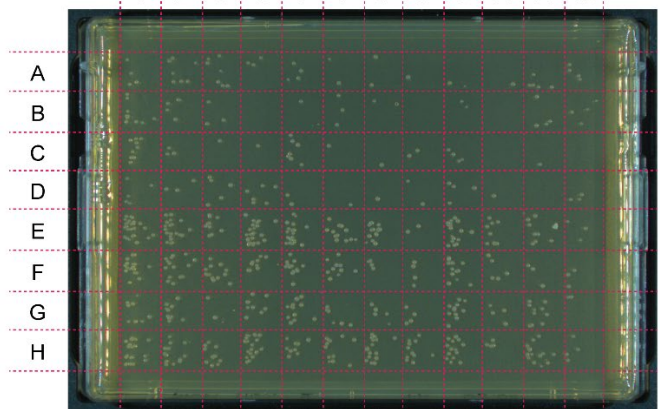

**Figure 4 | Workflow for high-throughput transformation of Golden Gate reactions. (A)** Workflow visualization of the individual steps described in 3.3. **(B)** Results of transformation plating by 12-well drop gravity flow using a 12-channel pipette. The top panel shows the plain transformation plate; the bottom panel highlights the individual lanes corresponding to the respective wells of the transformation plate. 20 µL of each transformation mixture is carefully

transferred to a 120 x 120 mm petri dish, prearranged at an angle of ~30°. The drops create a gradient of the transformation mixture by gravity flow. A representative plate is shown after overnight incubation at 37°C. **(C)** Results of 96-well transformation spotting using a 96-well pin pad-based robot. The top panel shows the plain transformation plate; the bottom panel highlights the individual lanes and rows corresponding to the respective wells of the transformation plate. The pin pad based robot transfers droplets in a 7 x 7 grid and visits the transformation mixture only once per row to create a dilution series. A representative plate is shown after overnight incubation at 37°C.

1. Prepare the appropriate agar medium with the required antibiotic. In this case, LB 2% agar supplemented with kanamycin as a selection marker is used (*see* **Note 27**).
2. Carefully pour the agar plates, avoiding the formation of bubbles. Do not move plates until completely set to ensure a flat and even surface. For drop plating, use 120 x 120 mm plates are used (Fig. 4B). For 96-well plating with a pin-based screening robot, use SBS format or other appropriate consumables (Fig. 4C) (*see* **Note 28, 29**).
3. Make sure the plates are properly dried but not overdried (*see* **Note 30**).
4. Transformation procedure:
    (a) Add 20 µL of competent cells to the Golden Gate reactions and incubate for 30 min on ice (*see* **Note 31**).
    (b) Prepare a 96-well deep-well plate with 500 µL recovery medium and preheat the plate(s) to 37 °C (*see* **Note 32**).
    (c) Heat shock cells in a preheated 384-well PCR cycler with the required settings (*see* **Note 33**).
    (d) Transfer cells to the prepared 96-well deep-well plate and recover cells for 45-60 min at 37 °C with frequent shaking.
    (e) Transfer the 96-well deep-well plate to a centrifuge and pellet the cells at 3000 x *g* for 5 min. Remove 450 µL of media using a 12- or 96-channel multichannel pipette and gently resuspend the cells in the remaining liquid (*see* **Note 34**).
5. Two platting methods can be used to reliably plate high-throughput transformations at higher densities (Fig. 2B-C).

12-well drop gravity flow plating using multichannel pipettes (Fig. 4B):
    (a) Position a 120 x 120 mm plate at an angle of approximately 30°.
    (b) Carefully transfer 20 µL of the transformation mixture to the top of the tilted plate (*see* **Note 35**). By manually tilting the plate to an angle of up to 90°, the speed and run distance of the drops can be adjusted.
    (c) When the drops are sufficiently spread, place the plate on a flat bench to dry the liquid.
    (d) Plate one row of the transformation mixture from the 96-well plate per 120 x 120 mm plate.

       96-well transformation pinning (Fig. 4C):
         (a) Transfer transformations to a 384 microtiter plate.
         (b) Use the 96-pin pin-pad to spot transformations on SBS solid media plates with the appropriate selection media in a 7 x 7 grid.
         (c) It is important to visit the transformation mixture only once for each row of the 7 x 7 grid to obtain a miniaturized dilution series.

6. Incubate the transformation plates inverted at 37 °C overnight (*see* **Note 36**).
7. Document your transformation plates using an imaging system. In this case, fluorescent TUs are assembled they should be imaged under the appropriate fluorescence conditions and standard illumination settings.

### *3.4    cPCR validation of constructs for subsequent sequence validation*

Once DNA constructs have been made and transformed, they need to be validated. The method presented here has a direct visible readout based on the expressed fluorescent protein. However, for applications without direct readout, other validation procedures are required. Therefore, a method based on colony PCR is described (Fig. 5). The use of *E. coli* cells as the PCR template allows the process to be accelerated in contrast to restriction digest pattern analysis. The amplicon size provides information about the DNA assemblies and may allow identification of the problematic part(s). In general, a diagnostic method should be sufficient to confirm Golden Gate assemblies beyond the basic part construction. Sequencing of plasmids is only required for basic part construction or if methods have been used that potentially introduce mutations (e.g. PCR amplification steps). cPCR validation of Golden Gate assemblies takes approximately half a working day from start to finish.

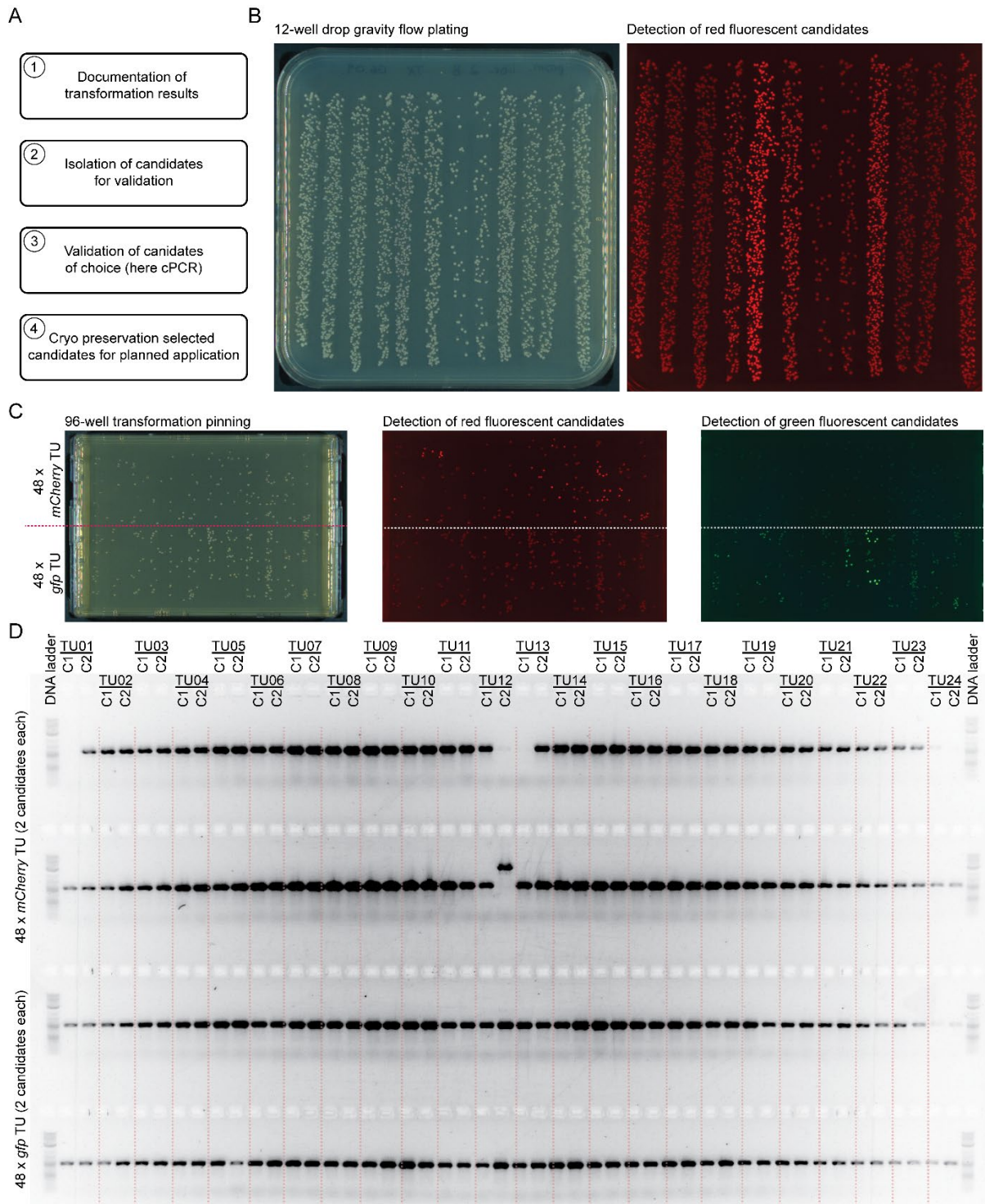

**Figure 5 | Workflow and example of construct validation. (A)** Workflow visualization of the individual steps described in 3.4. **(B)** Example image of 12-well drop gravity plating for the assembly of *mCherry* transcription units. Shown is a standard image (left) and an image under conditions that allow the detection of mCherry (right). Different fluorescence intensities can be observed at the colony level and are an indicator of the strength of promoters of unknown sequence. A representative result is shown. **(C)** Example images of 96-well plating using a pinning robot for the assembly of *mCherry* (top) and *gfp* (bottom) transcription units. Shown is a standard image (left) and images under conditions that allow detection of mCherry (middle) and GFP (right). Different fluorescence intensities can be

observed at the colony level and are an indicator of the strength of promoters of unknown sequence. A representative result is shown. **(D)** Two candidates from each transformation for the 48 x *mCherry* and 48 x *gfp* TU constructs were tested by cPCR to validate the TU constructs. Nearly all reactions show the expected band and at least one candidate for each TU construct appears to be correct. For visualization purposes, only the top row is labeled. The top half shows the 2 x 48 x *mCherry* assemblies and the bottom half shows the 2 x 48 x *gfp* assemblies. The respective *mCherry* and *gfp* positions correspond to the same promoter sequences. C1 = Candidate 1, C2 = Candidate 2, red dotted lines separate the different TU assemblies.

1. In the case of this protocol, functionality can be validated by imaging plates in the respective red and green fluorescent protein channels (Figure 5B-C).
2. Fluorescence does not necessarily confirm the correct assembly of candidates, so additional validation is performed by colony PCR (cPCR) (*see* **Note 37**).
3. Two colonies from each Golden Gate assembly are picked with a colony picking robot into a 384-well grid onto LB kanamycin agar plates. The plates are incubated at 37 °C for at least 6 hours.
4. 5 µL of PCR Master Mix (*see* **Note 38**) are dispensed into a 384-well PCR plate using a bulk dispenser (*see* **Note 39**). The composition of the PCR Master Mix is as follows:

|  | 1 x mix | 384 x mix (incl. 10% backup) |
|---|---|---|
| 2 x ready-to-load Master Mix | 2.5 µL | 1,056 µL |
| forward primer (10 µM) | 0.1 µL | 42.24 µL |
| reverse primer (10 µM) | 0.1 µL | 42.24 µL |
| ddH$_2$O | 2.3 µL | 971.52 µL |

5. Transfer cell material from the prepared agar plate to the PCR reactions using a 384-well pin pad.
6. Perform the colony PCR based on the polymerase and primer characteristics used. The following settings were used in this protocol:

    30 s 94 °C - [15 s 94 °C - 30 s 50 °C - 70 s 68 °C]$_{35 \text{ cycles}}$ - 5 min 68 °C - hold at 8 °C

7. Load 4 µL of the PCR reaction onto a 1% agarose gel and run at 100 V for 1 h (*see* **Note 40**). DNA loading dye is added directly to the gel.
8. Image and analyze the agarose gel using an appropriate documentation system (Fig. 5D).
9. Select suitable candidates from the arrayed agar plate used as cell material donor for the cPCR and inoculate overnight in appropriate medium for cryopreservation and downstream applications.

*3.5   Summary & Perspectives*

Acoustic dispensers have revolutionized molecular biology [19]. They allow highly parallelized reactions in nanoliter volumes. The transfer of liquids is free of consumables, making the use of acoustic dispensers an economical and sustainable strategy. Their versatility has been demonstrated in many different applications. We provide a detailed protocol for using acoustic dispensers for high-throughput Golden Gate assemblies. We also provide details on high-throughput transformation, selection and validation of DNA assemblies. We believe this technology will become the standard for molecular microbiology applications, and we have shown only one aspect of the use of acoustic dispensers.

4.   Notes

1. It is mandatory to perform efficiency tests for new acceptor plasmids, altered numbers or drastic fragment size changes prior to large-scale experiments to estimate DNA assembly efficiency.
2. It is recommended that each new batch of competent cells be tested for efficiency. Efficiency can be determined by transforming a standard plasmid of known concentration. CFUs can be used to estimate transformation efficiency in terms of CFUs per µg of plasmid DNA.
3. Efficiency testing is recommended when reagents are changed. This is especially true when new suppliers or in-house purified enzymes are used.
4. Consumables should be compatible with the screening robot used, it is recommended to use the supplier's single well plates or tested alternative consumables. In this chapter, the Rotor HDA+ from Singer Instruments has been used with the single well plates from the same manufacturer, which have been optimized with regard to their surface area.
5. Other formats will work as long as they are compatible in size with the multichannel pipettes used. If high-throughput plating is not used, standard petri dishes and glass bead-based spreading or alternative methods will also work.
6. 384-well plates are used as acceptor plates for the acoustic dispenser. Standard 96-well plates are too high for the acoustic dispensers used in this chapter. Low profile 96-well plates may be compatible with the acoustic dispenser. However, they must be skirted and a PCR cycler compatible with these plates must be available. For several reasons, we have adapted our protocol to use 384-well PCR plates as acceptor plates. Important: The dimensions of the 384-well PCR plate must be adapted to the operating system of the acoustic dispenser.
7. Any type of competent cells with sufficient transformation efficiency can be used. Transformation can also be performed by electroporation. However, this method may not be compatible with high throughput. It is strongly recommended to test the efficiency of batches of competent cells prior to large-scale experiments. Assembly efficiencies should be tested with varying numbers of fragments to ensure that the correct volume is plated after transformation.
8. Wild type strain used in this study. Any other experimental model strain is also suitable.
9. This strain is required to propagate the *ccdB* containing acceptor plasmids used. CcdB is a gyrase inhibitor and causes cell death in strains without CcdA antitoxin or gyrase mutations (e.g. *gyrA462*).
10. Standard cloning strain that is highly sensitive to CcdB. However, any other cloning strain or wild-type strain will work. CcdB sensitivity is advantageous to minimize the risk of cells containing only the acceptor plasmid. New strains can be easily tested by transforming a *ccdB*-encoding plasmid, a positive control (e.g. pUC-based plasmid) must be included in this experiment to ensure that the cells are competent.
11. Any plasmid purification method is suitable. However, for large-scale purification of plasmid DNA protocols using magnetic beads are preferred. Suitable open source protocols are for example available at www.bomb.bio [23] and are used here and in previous studies in combination with the acoustic dispenser [11].
12. For annealing, 4.5 µL each of 100 µM primer and 1 µL 10X annealing buffer (10X annealing buffer: 10 mM Tris(-HCl), pH 7.5-8.0, 50 mM NaCl and 1 mM EDTA) in a PCR reaction tube. The mixture is transferred to a PCR cycler. The sample is heated to 98 °C for 30 s and cooled to 8 °C at the lowest possible ramp speed of the instrument. The sample is then diluted to 20 fmol/µL and used immediately or stored at -20 °C. After multiple freeze-thaw cycles and an observed decrease in efficiency, a new annealing reaction should be performed. Important: If multiple oligonucleotides are fused, the DNA must be phosphorylated with e.g. T4 Polynucleotide Kinase according to the manufacturer's instructions prior to annealing.
13. Determination of DNA concentration using the absorbance ratio of $A_{260/280}$ is usually sufficient. However, if DNA assembly is inefficient, determination of DNA concentration using fluorescent dyes may be helpful in troubleshooting.
14. As an estimate, 40 ng of a 3 kb plasmid is equivalent to 20 fmol.
15. Carefully handle the plates to avoid cross-contamination when removing the seal. Before removing the seal, spin the plate briefly to ensure that the liquid is at the bottom of the wells.
16. Even when carefully sealed evaporation may be observed during prolonged freezer storage.
17. It is recommended to use multiple source plates. A master plate for enzymes and buffers and the appropriate DNA collection plates. Importantly, some acoustic dispenser models (e.g. Echo650T used in this chapter) are capable of using tubes in racks. The tubes can be arranged according to the user's needs and are particularly useful for avoiding multiple freeze-thaw cycles of valuable reagents.

18. Several tools have been developed for the automated generation of pick lists, many of which are available as open source on GitHub (e.g. from Henrike Niederholtmeyer's group: https://github.com/HN-lab/PyEcho).
19. This pick list is based on the Echo525 acoustic dispenser. Depending on the machine type, the pick list may differ in terms of the source plate type, for example. However, the general concept of defining the source well, destination well and transfer volume remains the same.
20. Carefully remove the seal, ideally while the reagents are still frozen, to minimize the risk of cross-contamination. It may be necessary to pulse spin the plate prior to seal removal.
21. Carefully inspect the source plate and repeat this step if necessary. Do not spin at too high x *g* to avoid damaging the source plate.
22. Operation may vary from device to device. In this protocol, the Echo525 or Echo650T (Labcyte) acoustic dispensers were used and the differences are indicated.
23. This plate definition is defined by the user based on the 384-well PCR plate consumables used. Any type of 384-well PCR plate can be used as long as the internal gripper can hold the plate and the plate is rigid enough not to bend. Importantly, 384-well plates that deviate from the standard square format may not be properly gripped and may fall out of the gripper during inverting.
24. Possible causes of exception errors are too low volume (outside of 15-65 µL range) or volume cannot be calculated (e.g. air bubbles, plate too cold). Exception errors may indicate that a particular reaction will not work because one or more reagents have not been transferred.
25. It is recommended to optimize the settings for each type of Golden Gate assembly (e.g., in regard to number of fragments and/or acceptor vector) to optimize the efficiency for optimal large-scale experiments.
26. However, if the assembly complexity is changed, it is always recommended to perform an initial small-scale test to obtain information on assembly efficiency.
27. Antibiotics and supplements in the media may be heat sensitive. Do not add additional components until the media has cooled to at least 60°C.
28. In this procedure, a Rotor HDA+ is used to inoculate the transformation mixture onto agar plates. It is recommended to use Singer PlusPlates in combination with this machine. However, any other pinning robot may be suitable for this process. However, it is important to test different consumables for optimal results.
29. SBS plates are ideally prepared with 35 to 50 mL of medium using either a 50 mL conical tube or an appropriate pipette. Carefully transfer the warm medium, avoiding air bubbles. Any air bubbles that are present can be either punctured or pushed to the edge with a sterile pipette tip as long as the agar has not solidified.
30. For best results, pour the plates the day before the assay and store the plates upside down at room temperature after the agar has solidified. The storage location should be away from direct sunlight, heat sources, or constant air flow (e.g., air conditioning). If plates are not to be used within 24 hours, it is recommended that the stack of plates be carefully wrapped in cling film or placed in a plastic bag and stored in the refrigerator.

31. In this experiment, competent cells are transferred from a reservoir (stored on ice) using a 12-channel multichannel pipette or 96-well pipetting station. Ensure proper mixing of DNA and cells by gently pipetting the transformation mixture up and down.
32. LB medium is used here. However, richer media or specific recovery media (e.g., SOC) may be used if transformation results in low colony numbers.
33. This protocol uses in-house prepared RbCl competent *E. coli* cells with the following heat shock settings 45 s 42 °C
34. If efficiencies are high, cell concentration may not be necessary.
35. Ensure that the plates have a flat surface without bubbles and are dry. Otherwise, lanes may merge and reactions may be cross-contaminated.
36. If no standard *E. coli* strain or temperature sensitive DNA constructs are used, the settings must be adjusted accordingly, e.g. incubation temperature and/or incubation time.
37. It is highly recommended to aim for an amplicon size of 500 to 1,500 bps and ideally, positive and negative candidates will result in amplicons of different sizes.
38. To prepare the correct amount of PCR Master Mix, it is recommended to include 10 to 15% excess.
39. In this case, a ready-to-load 2x Master Mix was used to speed up the sample analysis by gel electrophoresis. However, any DNA polymerase is suitable for this experiment and the most economical was used.
40. The settings must be adapted to the available gel electrophoresis equipment.

**Acknowledgments**

This work was supported by the Max Planck Society within the framework of the MaxGENESYS project (DS) and the European Union (NextGenerationEU) via the European Regional Development Fund (ERDF) by the state Hesse within the project "biotechnological production of reactive peptides from waste streams as lead structures for drug development" (DS). We are grateful to all laboratory members for extensive discussions on the cloning and validation steps of Golden Gate assemblies. We thank Ehmad Chehrghani Bozcheloe for the visualization of the acoustic dispenser. All materials are available on request from the corresponding author.